\documentclass[journal]{IEEEtran}
\ifCLASSINFOpdf
\else
\fi
\usepackage{amsmath}
\usepackage{amssymb}
\usepackage{amsthm}
\usepackage{bm}
\usepackage[numbers,sort&compress]{natbib}
\usepackage{graphicx}
\usepackage{epstopdf}
\usepackage{epsfig}
\usepackage[pdftex,CJKbookmarks = true,  bookmarksnumbered, bookmarksopen,  colorlinks,linkcolor = black, anchorcolor = black, citecolor = black, linktocpage = true]{hyperref}

\begin{document}
%
\title{Low PMEPR OFDM radar waveform design using the iterative least squares algorithm}
%
%
%

\author{Tianyao~Huang and Tong Zhao

\thanks{Part of this work \cite{Huang2015} was submitted to IET Radar Conference 2015.}
\thanks{T. Huang and T. Zhao are both with Radar Research (Beijing), Leihua Electronic Technology Institute, AVIC and also with Aviation Key Laboratory of Science and Technology on AISSS.}
\thanks{Email:\{huangty,zhaotong\}@ravic.cn.}
}

\maketitle

\begin{abstract}
This letter considers waveform design of orthogonal frequency division multiplexing (OFDM) signal for radar applications, and aims at mitigating the envelope fluctuation in OFDM. A novel method is proposed to reduce the peak-to-mean envelope power ratio (PMEPR), which is commonly used to evaluate the fluctuation. The proposed method is based on the tone reservation approach, in which some bits or subcarriers of OFDM are allocated for decreasing PMEPR. We introduce the coefficient of variation of envelopes (CVE) as the cost function for waveform optimization, and develop an iterative least squares algorithm. Minimizing CVE leads to distinct PMEPR reduction, and it is guaranteed that the cost function monotonically decreases by applying the iterative algorithm. Simulations demonstrate that the envelope is significantly smoothed by the proposed method.
\end{abstract}



%
\IEEEpeerreviewmaketitle

\section{Introduction}
As a multi-carrier modulation waveform, OFDM is employed both in communication and radar applications. It has received growing attention from radar community for its excellent performance since the pioneer work on multi-carrier phase-coded (MCPC) signal by Levanon \cite{Levanon2004} (Chapter 11).
OFDM is regarded as a low interception probability (LPI) waveform \cite{Lynch2009} (Chapter 5) for its instantaneous wideband property.
The frequency diversity in OFDM provides additional information on observed targets, and thus increases detection probabilities of the targets \cite{Sen2014}.
The wideband waveform mitigates possible fading and resolves the different multipath reflections \cite{Sen2009a,Sen2011}. In addition, due to advantages both in radar and communication applications, OFDM is also investigated for joint radar-communication systems \cite{sturm2011,Sit2014}.
\par
A severe drawback of OFDM is the envelope fluctuation problem. Superposition of different subcarriers generally brings notable variations in amplitude. The variation in instantaneous power leads to a large power backoff and reduces the efficiency of the amplifier, or results in nonlinear distortion. This letter focuses on mitigating the envelope fluctuations.
\par
The peak-to-mean envelope power ratio (PMEPR) or peak-to-average power ratio (PAPR) is usually employed to measure envelope fluctuations. In order to alleviate power back off or signal deformation, many approaches have been proposed to reduce PMEPR/PAPR.
\par
Some certain amplitude-phase modulations to generate low-PMEPR/PAPR OFDM waveforms are investigated \cite{Levanon2004,Rahmatallah2013,Narahashi1994}. Golay complementary sequences yield a signal of PAPR bounded with 2 \cite{Rahmatallah2013}.
Newman and Narahashi-Nojima phasing weights produce PMEPRs around 1.8 \cite{Levanon2004}. Schroeder method considers both amplitude and phase weight \cite{Levanon2004}. These solutions have closed-form expressions and are easy to apply, however, the resulting PMEPRs are not close enough to the achievable optimum \cite{Levanon2004}. In addition, since the modulations are immutable, the flexibility of these waveforms are restricted.
\par
Communication community has developed different approaches for PAPR reduction, which allow arbitrary digital data to be transmitted. Comprehensive surveys can be found in \cite{Banelli2014,Rahmatallah2013,Han2005,Jiang2008}. These algorithms are mainly divided into two categories: distortionless and distorted methods.
Compared with distorted approaches like clipping and companding transform techniques, distortionless methods avoid additional in-band and out-of-band distortion to the signals, preserve the orthogonality between subcarriers, and make it possible to independently process each subcarrier of radar returns.
\par
Among these distortionless approaches, tone reservation (TR) algorithm draws special interests. In a TR algorithm, some bits or subchannels are used to carry arbitrary information, which provides flexibility for radar waveform selection, and the rest are reserved to smooth the fluctuated envelope.
\par
This letter presents a novel TR algorithm to reduce the PMEPR of an OFDM signal.
The cost function and the solution are different from the standard TR method \cite{Rahmatallah2013}. In the standard TR approach, only the peak of the envelope is minimized. Note that the envelope fluctuation relies on overall extension of the waveform rather than a single sample. We choose the variance of the envelope as the penalty function, which considers both peaks and valleys of the envelope. And an efficient iterative least squares algorithm is proposed correspondingly. Theoretical analyses and numerical experiments demonstrate the effectiveness of the proposed algorithm.
\par
The rest of this letter is structured as follows. In Section \ref{sec:signal_model}, the signal model of OFDM waveform and the definitions of PAPR and PMEPR are introduced. Section \ref{sec:solution} models PMEPR reduction as an optimization problem and an iterative algorithm is developed. In Section \ref{sec:sim}, simulation results are presented to examine the performance of the proposed algorithm. Section \ref{sec:conclusion} briefly concludes this paper.
\section{OFDM Signal Model}\label{sec:signal_model}
Consider a complex-valued OFDM waveform containing $N$ carriers transmitted simultaneously with a frequency interval of $\Delta f$. The frequency of the $n$th carrier $f_n$ is $n \Delta f$, $n = 0,1,\dots,N - 1$. Each carrier contains $M$ amplitude-phase modulated bits with a duration of $t_b$. The bit duration $t_b$ is set as $1/\Delta f$, which yields the orthogonality between bits in different carriers. The pulse width $T_p$ is $Mt_b$. Denote the $m$th code in the $n$th carrier as $a_{n,m}$, $m = 0,1,\dots,M-1$, and the superposition of all carriers is
\begin{equation}
g(t) = \sum \limits_{n = 0}^{N-1} \sum \limits_{m = 0}^{M-1}a_{n,m} {\rm rect}\left({t-mt_b}\right) e^{ j2\pi f_n (t-mt_b)},
\end{equation}
where $\text{rect}(\cdot)$ is a rectangle function defined as
\begin{equation}
\rm{rect}(t) = \left\{
\begin{array}{l}
1, 0 \leq t \leq t_b, \\
0, \rm{else}.
\end{array}
 \right.
\end{equation}
\par
OFDM waveform suffers from a shortcoming that the amplitude $|g(t)|$ varies along with time.
Note that in an actual radar system, real-valued signal $\varrho(t) = \text{Re} \left[ g(t) \right]$ is transmitted and amplitudes $|\varrho(t) |$ is of pratical interest.
To enhance the detection range, power amplifier is always contained in the radar signal generator. High peaks encountered in $|\varrho(t)|$ could drive the power amplifier into saturation, which produces nonlinear distortion and interferences among subcarriers \cite{Rahmatallah2013}.
To avoid operating the power amplifier into saturation, input backoff (IBO) is applied \cite{Rahmatallah2013}. However, IBO shortens the detection range since the average power of the signal is decreased. It is preferable to reduce the envelope variations of the waveform, which remains the signal power unchanged and relieves  the signal distortion simultaneously.
\par
For a real-valued waveform $\varrho(t)$, the peak-to-average power ratio (PAPR), defined in (\ref{equ:PAPR}), is usually adopted to characterize the amplitude fluctuations.
\begin{equation}\label{equ:PAPR}
\text{PAPR}_{\varrho(t)} = \frac{\max_{t}|\varrho(t)|^2} {\frac{1}{Mt_b}\int_{0}^{Mt_b} |\varrho(t)|^2 dt}.
\end{equation}
The discrete-time PAPR is usually easier to handle than the continuous-time PAPR. Assume the sampling rate is $f_s = O_sN\Delta f$. When $O_s \geq 4$, the discrete-time PAPR well approximates the continuous-time counterpart \cite{Banelli2014}.
\par
For the convenience of using complex-valued waveform, PMEPR is selected instead of PAPR. And in low-bandwidth cases (i.e. the bandwidth is far less than the radio frequency), minimizing PMEPR is a good approximation to finding lower PAPR \cite{Levanon2004} (Chapter 11).
Denote the sampled waveform as $x_l = g(l/f_s)$, $l = 0,1,\dots,O_sNM -1$, and the discrete-time PMEPR is expressed as
\begin{equation}\label{equ:PMEPR_discrete}
\text{PMEPR}_{x_l} = \frac{\max_{l}|x_l|^2}{ E_l[|x_l|^2] },
\end{equation}
where $E_l\left[|x_l|^2\right] = \frac{1}{O_sNM}\sum_{l=0}^{O_sNM-1} |x_l|^2$ represents the mean of $|x_l|^2$.
Let $l = k+mO_sN$, $k = 0,1,\dots, O_sN -1$, and $x_l$ can be rewritten as
\begin{equation}\label{equ:wave_digital}
x_l= x_{k+mO_sN} = \sum_{n = 0}^{N-1} a_{n,m} e^{ j2\pi \frac{n}{O_sN}k}.
\end{equation}
Denote ${\bm x}_{m} = [ x_{mO_sN}, x_{1+mO_sN}, \dots,x_{O_sN -1 +mO_sN} ]^T \in \mathbb{C}^{O_sN}$. The signal within the $m$th bit can be presented by
\begin{equation}
{\bm x}_{m} = {\bm F} {\bm a}_m,
\end{equation}
where ${\bm a}_{m} = [a_{0,m}, a_{1,m},\dots,a_{N-1,m}]^T \in \mathbb{C}^{N}$ corresponds to the modulation symbols in the $m$th bit, and $\bm F \in \mathbb{C}^{O_sN \times N}$ is a Fourier transform matrix with $F_{kn} = e^{j2\pi\frac{n}{O_sN}k}$ as its $k$th row, $n$th column element.
Let ${\bm x} = [ {\bm x}_{0}^T,  {\bm x}_{1}^T,  \dots, {\bm x}_{M-1}^T]^T \in \mathbb{C}^{O_sNM}$ and ${\bm a} = [ {\bm a}_{0}^T,  {\bm a}_{1}^T,  \dots, {\bm a}_{N-1}^T]^T \in \mathbb{C}^{NM}$, then
\begin{equation}
{\bm x} = {\bm A} {\bm a},
\end{equation}
where ${\bm A} = \text{diag}(\bm F,\bm F,\dots,\bm F)\in \mathbb{C}^{O_sNM \times NM}$.
The goal of this paper is to design codes $\bm a$ which smooth $|\bm x|$.
\section{Problem Formulation}\label{sec:solution}
A TR algorithm divides $\bm a$ into two sub-vectors, $\bm a_{S^I}$ and $\bm a_{S^R}$, containing informative symbols and reserved symbols, respectively. Sets $S^I$ and $S^R$ are disjoint subsets of $S^{IR} = S^{I} \bigcup S^{R} = \{0,1,\dots,NM-1\}$. Symbols in $\bm a_{S^I}$ can be arbitrary. Given $\bm a_{S^I}$, symbols in $\bm a_{S^R}$ are reserved to mitigate envelope fluctuations.
\par
Informative bits can be carefully designed for some special purpose. For example, in a joint radar-communication system these bits bear communication information. In other cases without specific constraints, $\bm a_{S^I}$ can also be randomly chosen to increase the diversity of the waveform.
\par
Note that in a radar, energy of both $\bm a_{S^I}$ and $\bm a_{S^R}$ are useful and important. The performance of target detection mainly depends on the total energy of the transmitted $\bm a_{S^I}$ and $\bm a_{S^R}$. Echoes of both symbols are received and processed by the radar, and equally contribute to target detection. TR method allocates energy over carriers, but does not directly affect the total energy of transmission.
\par
A desirable criterion is to minimize PMEPR, i.e.
\begin{equation}\label{equ:opt_PMEPR}
\min_{\bm a_{S^R}} \frac{\max_l |x_l|^2}{E_l\left[ |x_l|^2 \right]}, \ s.t. \ \bm x = \bm A_{S^I} \bm a_{S^I} + \bm A_{S^R} \bm a_{S^R},
\end{equation}
where $\bm A_{S}$ denotes columns of $\bm A$ indexed in the set $S$. To abbreviate expressions, let $\beta = E_l\left[ |x_l| \right]$, $\bm c = \bm A_{S^I} \bm a_{S^I}$, $\bm b = \bm a_{S^R}$ and $\bm B = \bm A_{S^R}$. Due to the denominator of the objective function, (\ref{equ:opt_PMEPR}) is rather difficult to solve.
\par
A standard tone reservation approach \cite{Rahmatallah2013} simplifies (\ref{equ:opt_PMEPR}) by ignoring the denominator, which yields
\begin{equation}\label{equ:opt_PMEPR2}
\min_{\bm b} \|\bm x\|_{\infty}, \ s.t. \ \bm x = \bm c + \bm B \bm b,
\end{equation}
where $\| \bm x \|_{\infty} = \max_l |x_l|$ is the infinity norm of $\bm x$. Linear programming solves (\ref{equ:opt_PMEPR2}). A variation of (\ref{equ:opt_PMEPR2}) minimizes $E_l\left[ |x_l|^4 \right]$ instead \cite{Behravan2009}.
Both criteria seek to decrease the peak values in $|\bm x|$. However, the valley values of $|\bm x|$, which also contribute to envelop variations, are not increased by these two criteria. It is not guaranteed that a sequence $\bm b$ leading to a constant envelope is the optimizer of the criteria.
\par
This letter proposes a novel criterion which minimizes the variance of envelopes, and both peak and valley values are taken into account. We introduce \emph{coefficient of variation of envelope} (CVE) to evaluate the fluctuation degree of $\bm x$, which is defined as
\begin{equation}\label{equ:CV_discrete}
\text{CVE}_{x_l} =\frac{  E_l \left[ \left( \left|x_l\right|-E_p\left[|x_p|\right]\right)^2 \right] }
{\left(E_l\left[|x_l|\right]\right)^2}.
\end{equation}
CVE is closely related to PMEPR. It is satisfied that
\begin{equation}\label{equ:PMEPR_CEV}
1 \leq \sqrt{\text{PMEPR}_{x_l}} \leq \sqrt{O_sNM}\sqrt{\text{CVE}_{x_l}} + 1,
\end{equation}
where the equalities holds if and only if $\bm x$ has a constant envelope; see proofs in Appendix \ref{app:proof}. When the envelope is not constant and the number of samples $O_sNM$ are large, bounds in (\ref{equ:PMEPR_CEV}) could be loose. However, it still can be indicated from (\ref{equ:PMEPR_CEV}) that 1) codes in $\bm b$ which result in a constant envelope, if the codes exist, lead to the minimum of CVE, and 2) a small PMEPR is guaranteed if CVE is small.
Omit the denominator in (\ref{equ:CV_discrete}), optimization problem becomes
\begin{equation}
\min_{\bm b} E_l \left[{\left|x_l\right|^2} - {\left(E_p\left[ |x_p| \right]\right)^2}\right]
, \ s.t.\  \bm x = \bm c + \bm B \bm b,
\end{equation}
or equivalently,
\begin{equation}\label{equ:CV_discrete2}
\min_{\bm b}  \sum\limits_{l=0}^{O_sNM-1}\left( \left|x_l\right|-E_p\left[ |x_p| \right]\right)^2
, \ s.t.\  \bm x = \bm c + \bm B \bm b.
\end{equation}
It still holds that codes in $\bm b$ making the envelop constant are the optimal solution of (\ref{equ:CV_discrete2}).
Reformulate (\ref{equ:CV_discrete2}) in a matrix form
\begin{equation}\label{equ:opt}
\begin{split}
&\min_{\bm b}  \left(  \bm B \bm b + \bm c - \beta e^{j\bm \theta} \right)^{H}
\left(   \bm B \bm b + \bm c - \beta e^{j\bm \theta} \right),\\
&\ s.t. \
\beta = \frac{1}{O_sNM}\| \bm B \bm b + \bm c \|_1,
\bm \theta = \measuredangle ( \bm B \bm b + \bm c),
\end{split}
\end{equation}
where $\|\cdot\|_1$ denotes the $\ell_1$ norm of a vector, and $\measuredangle x$ is the angle of complex-valued $x$.
A closed-form solution to (\ref{equ:opt}) is hard to obtain due to involved complex and nonlinear operations.
\par
Symbols in $\bm b$ and its functions $\beta, \bm \theta$ are calculated alternately.
Assume that at the $i$th iteration, $\bm b^{(i)}$ is obtained, then
\begin{equation}\label{equ:beta_theta}
\begin{split}
\beta ^ {(i)}= \frac{1}{O_sNM}\| \bm B \bm b^ {(i)} + \bm c \|_1, \bm \theta ^ {(i)}= \measuredangle ( \bm B \bm b^ {(i)} + \bm c).
\end{split}
\end{equation}
Substitute $\beta = \beta^{(i)}$ and $\bm \theta = \bm \theta^{(i)}$ in (\ref{equ:beta_theta}) into (\ref{equ:opt}), and (\ref{equ:opt}) reduces to a least squares problem, of which the solution is
\begin{equation}\label{equ:CV_solution}
\bm b^{(i+1)} = -\bm B^{\dag}\left(\bm c - \beta^{(i)} e^{j\bm \theta^{(i)}}\right),
\end{equation}
where $\bm B^{\dag}$ is the pseudo inverse $\bm B^{\dag}= \left( \bm B^H \bm B\right)^{-1}\bm B^H$.
\par
It is proved that the cost function in (\ref{equ:opt}) is monotonically nonincreasing versus the iteration counter $i$; see Appendix \ref{app:conv}.
\par
{\it{Remarks}.}
Partitioning of the set $S^{IR}$ affects the achievable PMEPR level. Intuitively, increasing the number of free variables, i.e. $|S^R|$, benefit lowering the PMEPR. But in some applications, choice of $S^I$ are rigid. For example, in a jamming environment, jammed subcarriers are contained in $S^I$ and the corresponding bits are set zeroes to evade jammer. Without such purposes, division of $S^{IR}$ is arbitrary and has the potential to be optimized during the waveform design processing. We may remain further discussions as the future work.
\section{Numerical Results}\label{sec:sim}
Numerical experiments are performed to demonstrate the advantages of the proposed method over the standard TR approach (\ref{equ:opt_PMEPR2}) and its variation \cite{Behravan2009}.
To solve problem (\ref{equ:opt_PMEPR2}) and its variation \cite{Behravan2009} we apply CVX, a package for specifying and solving convex programs \cite{cvx,gb08}.
These three algorithms are denoted as 'TR-CVE', 'TR-max' and 'TR-E$|x|^4$', corresponding to their cost functions, respectively.
In the following experiments, the baseband bandwidth is $B = 50$ MHz, and the sampling rate $f_s = 500$ MHz.
\subsection{Convergence}\label{sub:conv}
\begin{figure}[!h]
\centering
\includegraphics[width=3in]{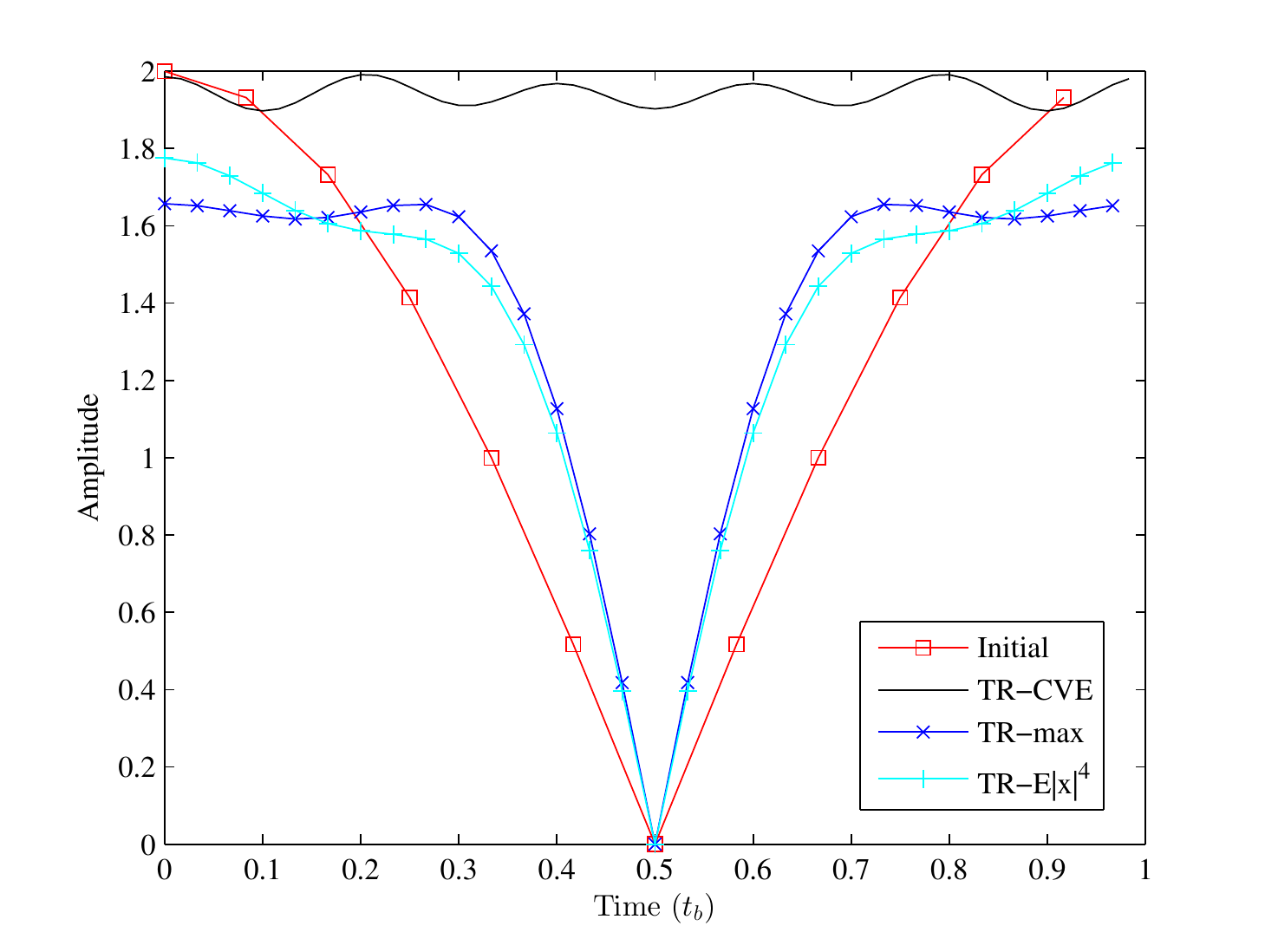}
\caption{Envelopes of the initial and optimized waveforms.}
\label{fig:evelopes}
\end{figure}

Suppose that there are $N = 6$ subcarriers, of which each contains $M = 1$ bit. Two subcarriers, $n =2,3$, are predefined with $a_{2,0} = a_{3,0} = 1$. The rest subcarriers are reserved as free variables for PMEPR reduction.
For the proposed method, set the initial value as $\bm b^{(0)} = \bm 0$, and alternately calculate $\beta$, $\bm \theta$ with (\ref{equ:beta_theta}) and the symbols $\bm b$ with (\ref{equ:CV_solution}), respectively.
After 800 iterations, the final symbols $\bm b$ are obtained.
Envelopes of the optimized waveforms and the initial counterpart are depicted in Fig. \ref{fig:evelopes}. Note that envelopes are not normalized, and the average powers are not 1,  $E_l[|x_l|^2] \neq 1 $.
\par
As shown in Fig. \ref{fig:evelopes}. The fluctuations in the envelope declines significantly using the proposed method. PMEPR decreases from 2 to 1.05. The standard tone reservation approach and its variation decrease peak values in $|x_l|$, but PMEPRs are slightly reduced to 1.30 and 1.51, respectively.
\par
\begin{figure}[!h]
\centering
\includegraphics[width=3in]{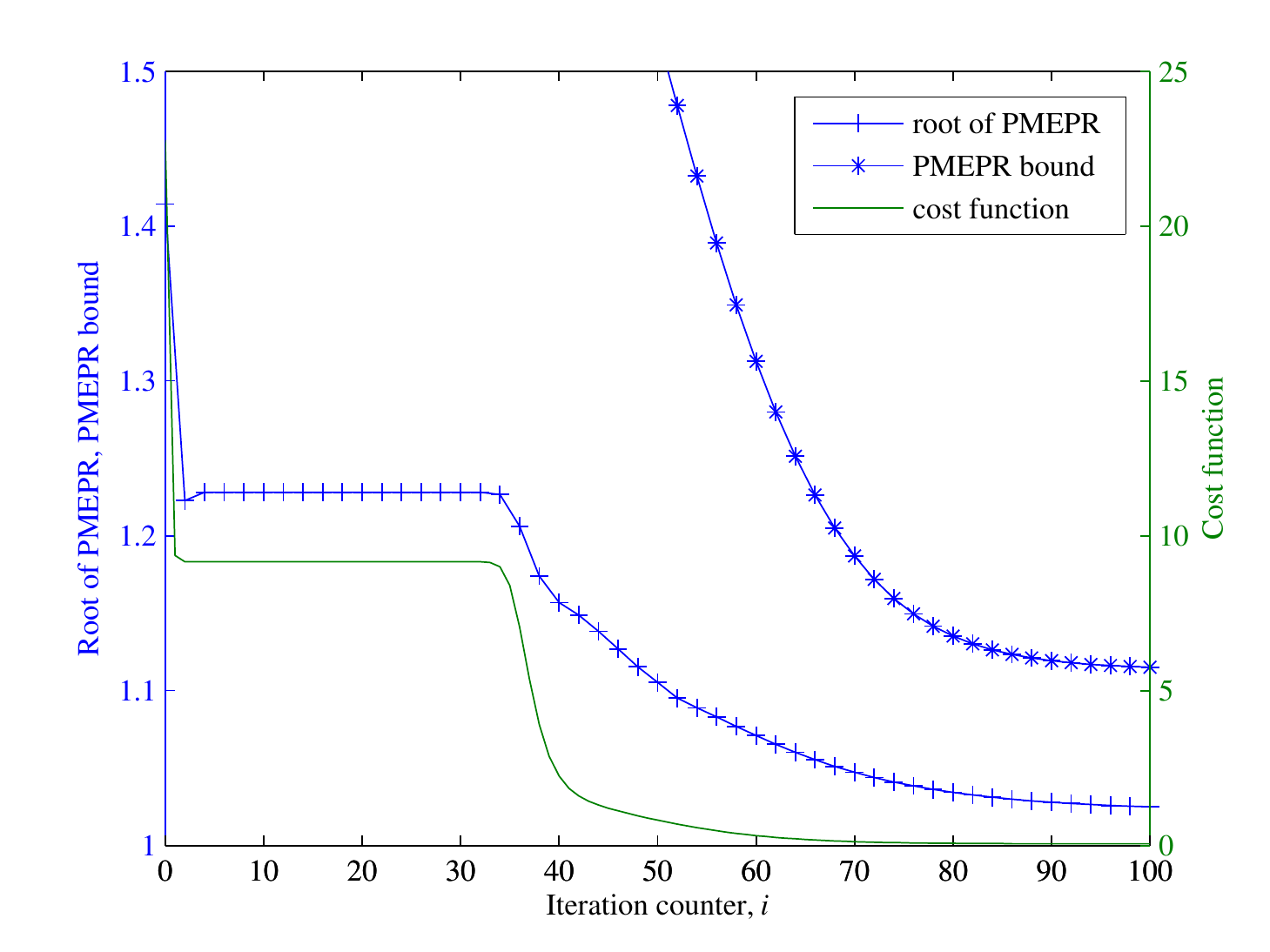}
\caption{Roots of PMEPRs, the upper bounds in (\ref{equ:PMEPR_CEV}) and the cost function in (\ref{equ:opt}) versus iteration number.}
\label{fig:convg}
\end{figure}

The roots of PMEPRs, the corresponding upper bound in (\ref{equ:PMEPR_CEV}) and
the cost function of (\ref{equ:opt}) versus the iteration number are calculated and shown in Fig. \ref{fig:convg}, which demonstrates the monotonicity of the objective function in (\ref{equ:opt}). PMEPRs are not monotonically reduced, since PMEPR is slightly different to the cost function. And the upper bound is not tight (values above 1.5 are not displayed). However, PMEPR, its upper bound and the cost function have similar trend. When the bound and the cost function are small, PMEPRs are also small.
\subsection{CCDF of PMEPR}
\begin{figure}[!h]
\centering
\includegraphics[width=3in]{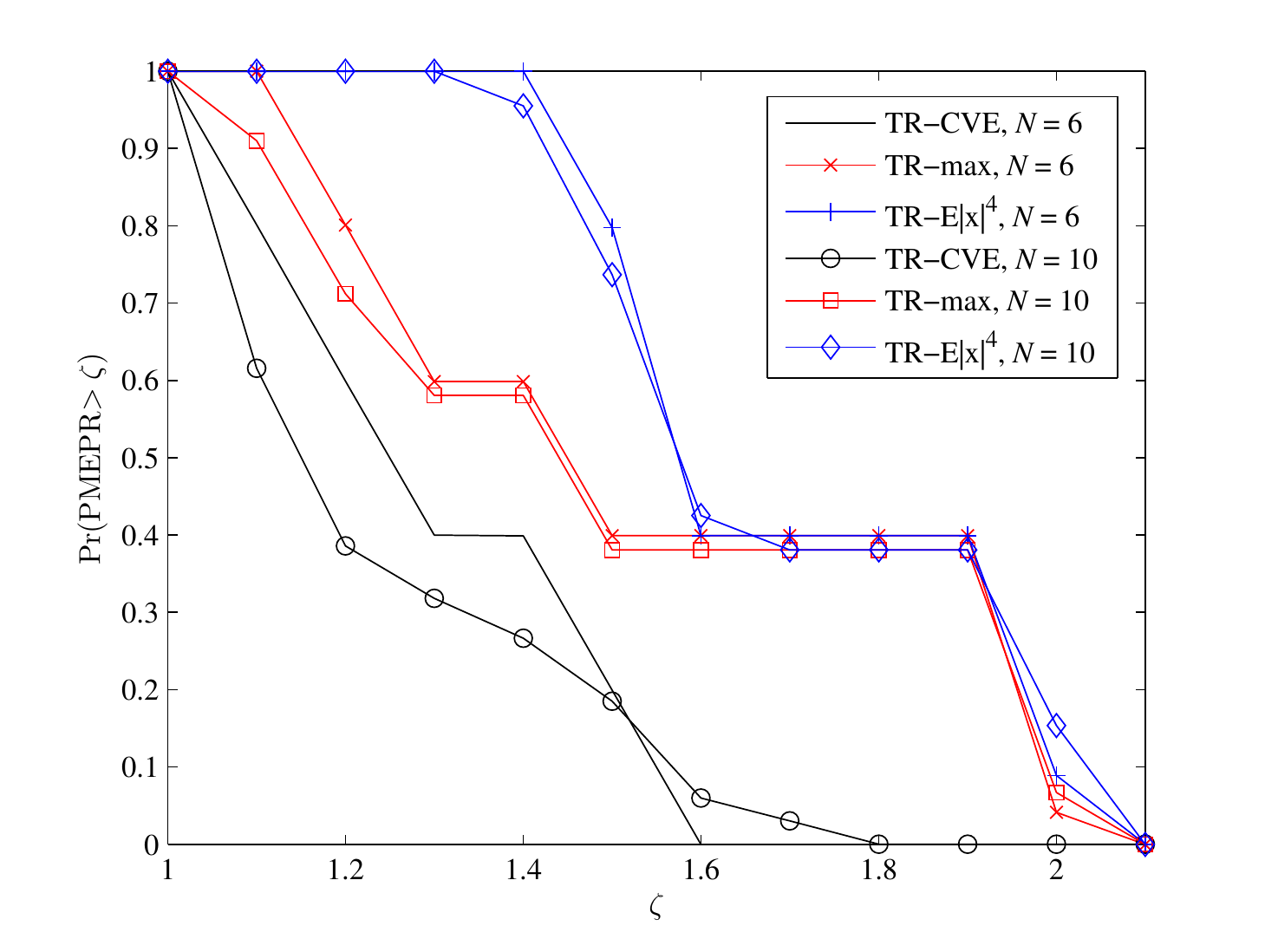}
\caption{CCDFs of PMEPRs for random informative symbols $\bm a_{S^I}$.}
\label{fig:ccdf}
\end{figure}

In this subsection, the informative subcarriers and symbols therein are randomly selected, and the complementary cumulative distribution function (CCDF) of PMEPR is examined. We consider two cases $N = 6$ and $N = 10$, among which two are randomly selected as informative subcarriers. Set $M = 10$, and the codes $a_{n,m} \in {\bm a}_{S^I}$ are equally likely valued among $\{-1,1,j,-j\}$.
The CCDFs of PMEPRs are displayed in Fig. \ref{fig:ccdf}, which is obtained with $5 \times 10^4$ Monte-Carlo trials. Fig. \ref{fig:ccdf} shows that the proposed iterative algorithm leads to lower PMEPR levels than the existing  approaches (\ref{equ:opt_PMEPR2}) and \cite{Behravan2009}.
\subsection{Ambiguity function and detection performance}
\begin{figure}[!h]
\centering
\includegraphics[width=3in]{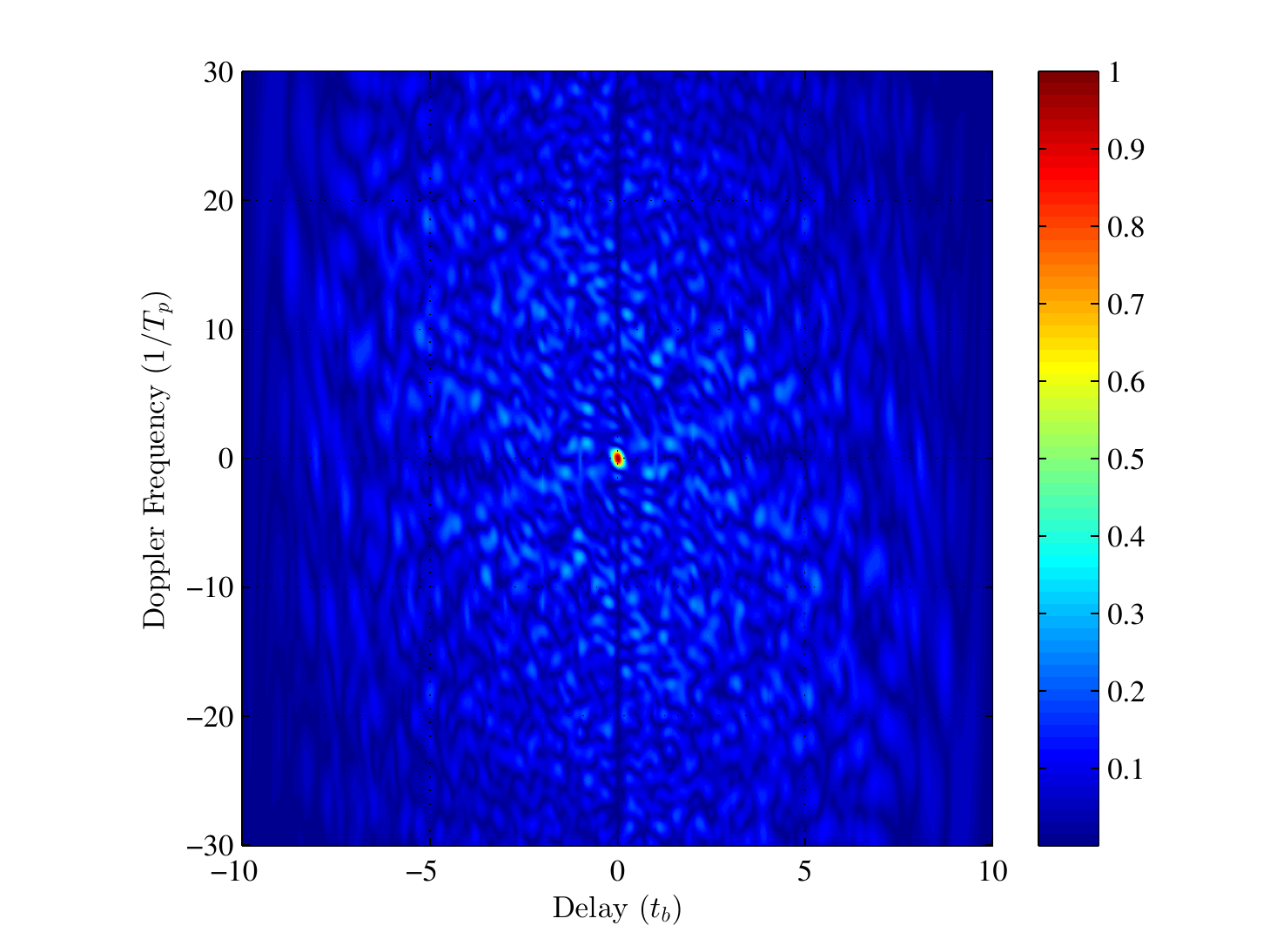}
\caption{Ambiguity function of the OFDM waveform produced with TR-CVE algorithm.}
\label{fig:AFLS}
\end{figure}

\begin{figure}[!h]
\centering
\includegraphics[width=3in]{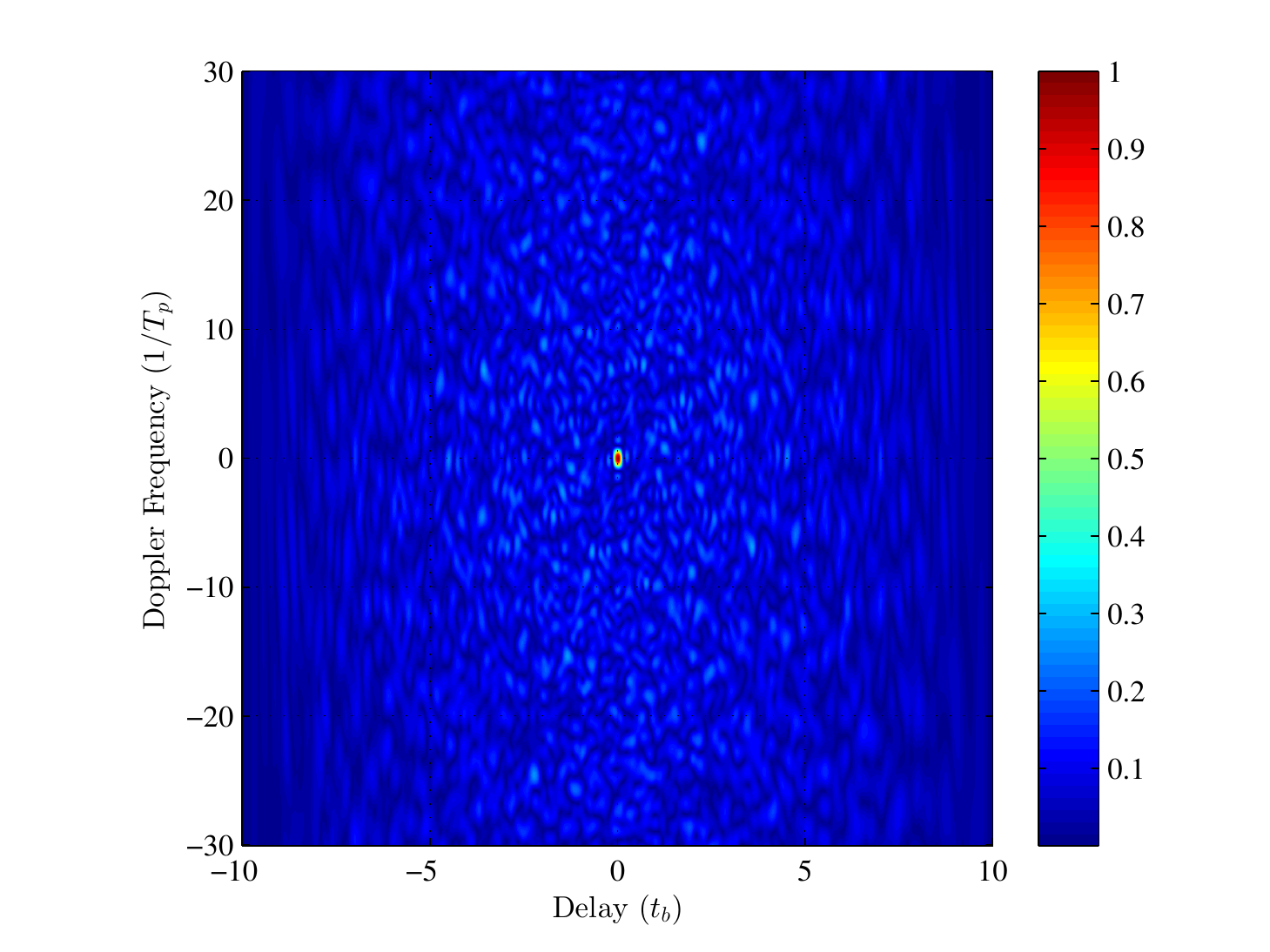}
\caption{Ambiguity function of the 'Uniform OFDM' waveform.}
\label{fig:AFuniform}
\end{figure}

In this subsection, the ambiguity functions and target detection performances of the optimized waveforms are discussed. There are $N = 6$ subcarriers with  $M = 10$ bits in each carrier. Two carriers, $n = 2,3$, are predefined, and Chu codes are chosen with its $m$th phase as $a_{n,m} = e^{j\gamma \frac{\pi}{M} m^2}, n = 2, 3$, where $\gamma = 1$ for $n = 2$ and $\gamma = -1$ for $n =3$. The rest bits are reserved as free variables for TR algorithms. We also examine 'Uniform OFDM' with the rest bits of a uniform amplitude but random phases $a_{n,m} = e^{j\phi_{n,m}}$,  $n = 0,1,4,5$, where $\phi_{n,m}$ is independently uniformly distributed over $[0,2\pi)$. The ambiguity functions are shown in Fig. \ref{fig:AFLS} $\sim$ Fig. \ref{fig:AFuniform}.
\par
\begin{figure}[!h]
\centering
\includegraphics[width=3in]{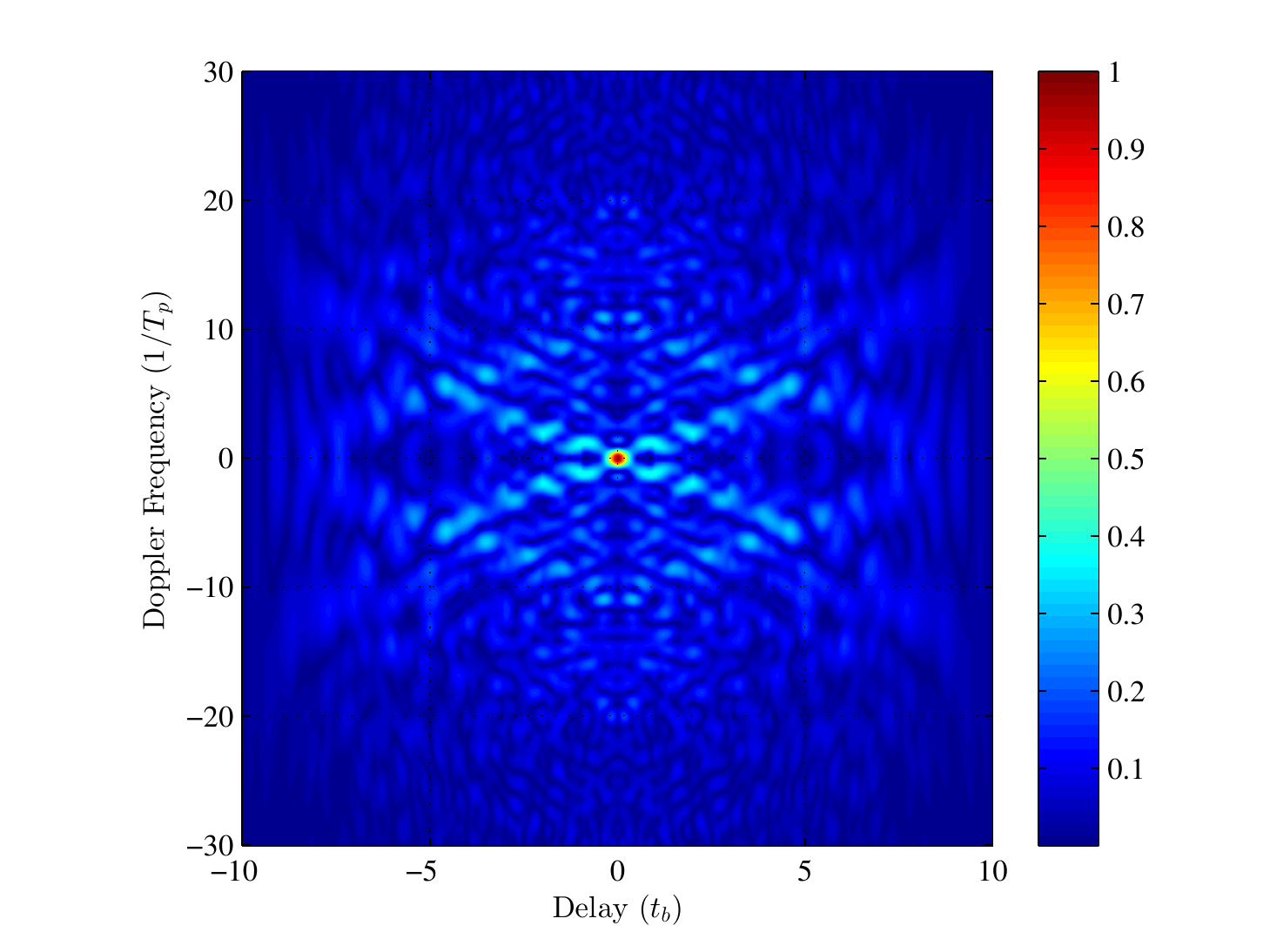}
\caption{Ambiguity function of the OFDM waveform produced with TR-max algorithm.}
\label{fig:AFcvx}
\end{figure}
\begin{figure}[!h]
\centering
\includegraphics[width=3in]{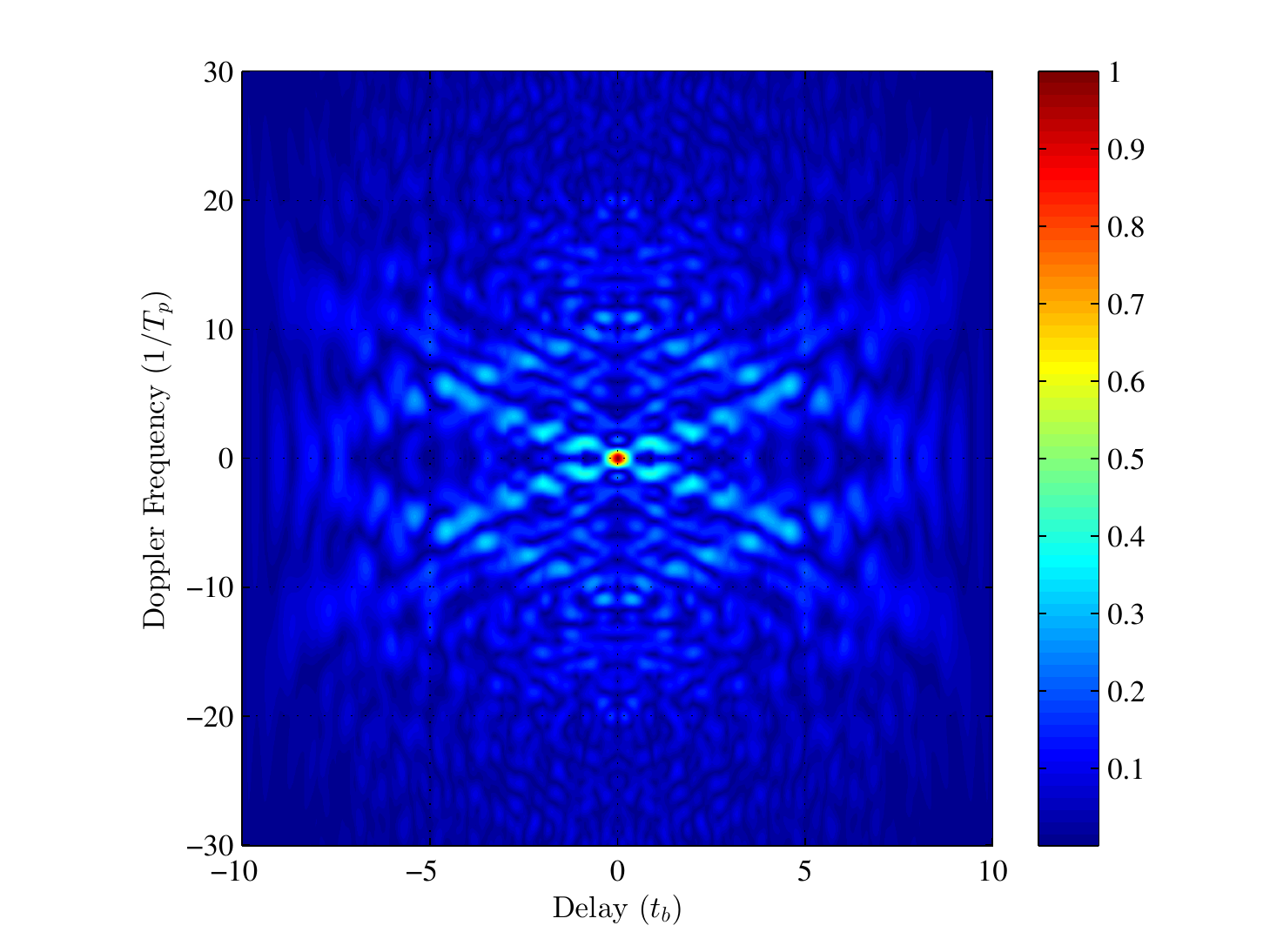}
\caption{Ambiguity function of the OFDM waveform produced with TR-E$|x|^4$ algorithm.}
\label{fig:AFcvxx4}
\end{figure}
As shown in these figures, waveform designed with TR-CVE algorithm has similar ambiguity function with that of 'Uniform OFDM' waveform. And waveforms by TR-max and TR-E$|x|^4$ have high sidelobes.
\par
\begin{figure}[!h]
\centering
\includegraphics[width=3in]{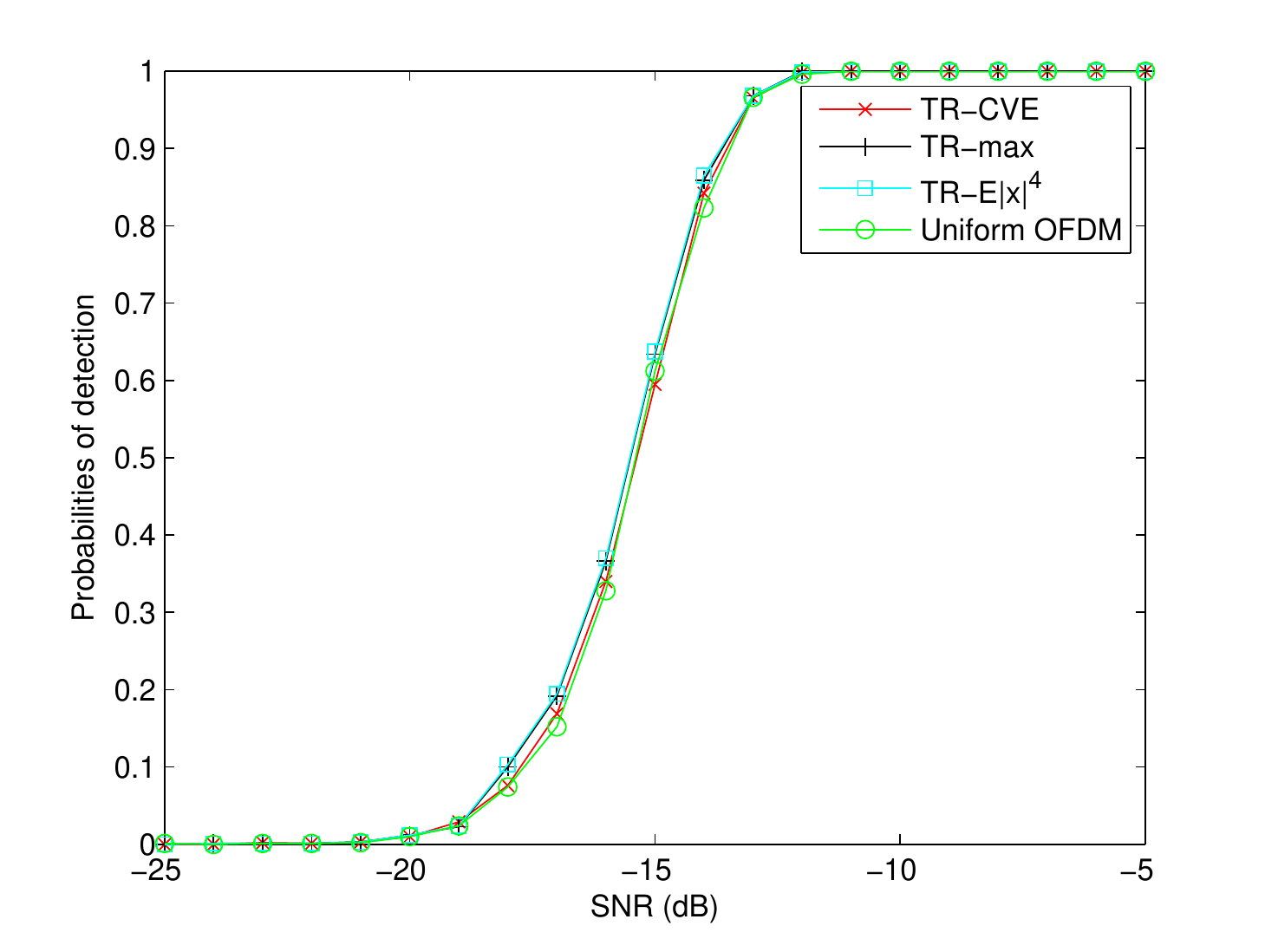}
\caption{Ambiguity function of the 'Uniform OFDM' waveform.}
\label{fig:pd}
\end{figure}

We also examine the detection performance of these waveforms. Waveforms are normalized such that the average power is unique. An ideal point target is considered. The noise power is set as $\sigma^2 = 1$. Noisy radar echoes are processed with matched filter. The results are compared with thresholds. For different OFDM waveforms, thresholds are selected such that the false alarm is $10^{-5}$. The radar cross section of the target varies to change the signal noise ratio (SNR). The detection probabilities of the target is shown in Fig. \ref{fig:pd}.
\par
As shown in Fig. \ref{fig:pd}, the detection probabilities of different waveforms are close to each other. 

\section{Conclusion}\label{sec:conclusion}
This letter considers a tone-reservation based algorithm to lower the PMEPR of an  OFDM waveform for radars. We propose coefficient of variation of envelopes (CVE) as a metric to measure the envelope fluctuation, and an iterative least squares method is devised consequently. Simulations show that the algorithm significantly improve the performance on PMEPR reduction.



\appendices
\section{}\label{app:proof}
Since $\max_{l} \left|x_l\right|^2 \geq E_l \left[ \left|x_l\right|^2 \right] \geq \left( E_l\left[\left|x_l\right|\right]\right)^2$, it holds
\begin{equation}
1 \leq \sqrt{\text{PMEPR}_{x_l}}
\le {\max_{l} \left|x_l\right| }/ { E_p\left[\left|x_p\right|\right] },
\end{equation}
\begin{equation}
\begin{split}
\frac{ \max_{l}\left|x_l\right| }{ E_p\left[\left|x_p\right|\right] }&-1
= \frac{ \max_{l}\left|x_l\right| - E_p\left[\left|x_p\right|\right]}{ E_p\left[\left|x_p\right| \right]}  \\
&\le {\sqrt{{\sum\limits_{l=0}^{O_sNM-1} \Big( \left|x_l\right|-E_p\left[\left|x_p\right|\right] \Big)^2}}}/ {E_p\left[\left|x_p\right|\right]}\\
&= \sqrt{O_sNM \cdot \text{CVE}_{x_l}},
\end{split}
\end{equation}
and then we have (\ref{equ:PMEPR_CEV}).
\section{}\label{app:conv}
Define the cost function in (\ref{equ:opt}) as $f(\bm b,\beta,\bm \theta) $. We prove
\begin{equation}
f\left(\bm b^{(i+1)},\beta^{(i+1)},\bm \theta ^{(i+1)}\right) \leq f\left(\bm b^{(i)},\beta^{(i)},\bm \theta ^{(i)}\right),
\end{equation}
where $\bm b^{(i)}$ is an arbitrary initial value and the rest variables are obtained with (\ref{equ:beta_theta}) and (\ref{equ:CV_solution}).
\begin{proof}
Since $\bm b^{(i+1)}$ is a solution of least squares, it obeys,
\begin{equation}
f\left(\bm b^{(i+1)},\beta^{(i)},\bm \theta ^{(i)}\right) \leq f\left(\bm b^{(i)},\beta^{(i)},\bm \theta ^{(i)}\right).
\end{equation}
Then we need prove
\begin{equation}\label{equ:conv_inequ}
f\left(\bm b^{(i+1)},\beta^{(i+1)},\bm \theta ^{(i+1)}\right) \leq f\left(\bm b^{(i+1)},\beta^{(i)},\bm \theta ^{(i)}\right).
\end{equation}
Define a cost function
\begin{equation}\label{equ:conv_opt}
\begin{split}
L\left(\beta,\bm \theta\right) &= {1}/{2} \left(  \bm y - \beta e^{j\bm \theta} \right)^{H}
\left(   \bm y  - \beta e^{j\bm \theta} \right),
\end{split}
\end{equation}
where $\bm y \in \mathbb{C}^N$ is a nonzero vector, $\beta \in \mathbb{R}$, $\bm \theta \in \mathbb{R}^N$, and each entry of $\bm \theta$ belongs to $[0,2\pi)$.
Set ${\bm y}=\bm B\bm b^{(i+1)} + \bm c$. To prove (\ref{equ:conv_inequ}), we only need to prove that the minimizer of (\ref{equ:conv_opt}) is (\ref{equ:beta_theta}).
\par
It can be found $\lim\limits_{\beta \to \infty}L\left(\beta,\bm \theta\right)=+\infty$, which indicates that the global minimizer of (\ref{equ:conv_opt}) is an extreme point.
Calculate the partial derivatives
\begin{equation}
\begin{split}
\frac{\partial L}{\partial \beta} = N\beta - \text{Re}\left( \bm y^H e^{j\bm \theta } \right),
\end{split}
\end{equation}
\begin{equation}
\begin{split}
\frac{\partial L}{\partial \bm \theta} = - \beta \text{Im}\left( e^{-j\bm \theta } \odot \bm y \right),
\end{split}
\end{equation}
where $\odot$ is element-wise multiplication.
The Hessian matrix
\begin{equation}
\bm H = \left[
\begin{array}{cc}
N & -\text{Im}\left( e^{-j\bm \theta^T } \odot \bm y^T \right)\\
-\text{Im}\left( e^{-j\bm \theta } \odot \bm y \right) & \beta \text{diag} \left( \text{Re}\left( e^{-j\bm \theta } \odot \bm y \right) \right)
\end{array}
\right].
\end{equation}
Solve $\frac{\partial L}{\partial \beta}= 0$ and $\frac{\partial L}{\partial \bm \theta}=\bm 0$, and the Hessian matrix becomes
\begin{equation}\label{equ:Hessian2}
{\bm H} |_{\frac{\partial L}{\partial \beta} = 0, \frac{\partial L}{\partial \bm \theta}=\bm 0} = \text{diag} \Big( [N, \pm \beta |y_0|,\dots,\pm \beta|y_{N-1}|] \Big).
\end{equation}
The Hessian matrix (\ref{equ:Hessian2}) is positive definite, if and only if
\begin{equation}\label{equ:solution}
\beta = \frac{1}{N}\| \bm y \|_1, \\
\bm \theta = \measuredangle \bm y,
\end{equation}
which yields ${\bm H}  = \text{diag} \Big( \left[N, \beta |\bm y|^{T}\right] \Big)$ and indicates that (\ref{equ:solution}), or (\ref{equ:beta_theta}), is the minimizer of (\ref{equ:conv_opt}).
\end{proof}
\bibliographystyle{IEEEtran}
\bibliography{OFDM_refs2}




\end{document}